\documentstyle [12pt]{article}
\advance\voffset by -1 in
\advance\hoffset by -.5 in
\def\z{\zeta}

\def\f{\phi}
\def\l{\lambda}
\def\m{\mu}

\def\p{\partial}

\def\O{\Omega}
\def\G{\Gamma}
\def\Om0{\Omega^0}
\def\O1{\Omega^1}

\def\hw{\hat{w}}

\def\res{{\rm res}}

\begin{document}
\centerline{{\bf On additional symmetries of the KP hierarchy}}
\centerline{{\bf and Sato's B\"acklund transformation}}

\vspace{.3in}
\centerline{{\bf L.A.Dickey}}

\vspace{.2in}
\centerline{Dept. Math., University of Oklahoma, Norman, OK 73019\footnote{
Internet: ldickey@nsfuvax.math.uoknor.edu}}

\small
\vspace{.5in}
\centerline{{\bf Abstract}}

\parindent .75in
\vspace {.2in}
\parbox{5in}{A short proof is given to the fact that the additional symmetries
of the KP hierarchy defined by their action on pseudodifferential operators,
according to Fuchssteiner-Chen-Lee-Lin-Orlov-Shulman, coincide with those
defined by their action on $\tau$-functions as Sato's B\"acklund
transformations. Also a new simple formula for the generator of additional
symmetries is presented. }
\normalsize
\parindent .25in

\vspace{.5in}
{\bf 0.} The so-called ``additional symmetries" of the KP hierarchy, i.e.,
symmetries which are not contained in the hierarchy itself and play such a
crucial role for the string equation independently appeared twice in remote
areas of the theory of integrable systems. On the one hand, they were
introduced
in works by Fuchssteiner, Chen, Lee and Lin, Orlov and Shulman, and others
who explicitly wrote the action of the additional symmetries on the
pseudodifferential operators and their wave functions. On the other hand, they
were found as B\"acklund transformations of $\tau$-functions by Sato and other
mathematicians of the Kyoto school. During a long period of time there was no
general evidence that these two kinds of symmetries coincide. This problem
received a great practical significance. As it was said, the symmetries of the
first kind are related to the string equation. They provide the Virasoro and
higher $W$-constraints. It is very important to know how do they act on
$\tau$-functions since the latter ones have a direct physical meaning as
partition functions in matrix models.

For lower additional symmetries the problem was solved in the positive sense by
direct though laborious calculations. The general proof that two types of
symmetries are, in fact, the same was given by Adler, Shiota and van Moerbeke
[1]. In our opinion, this proof had a little gap (it was tacitly assumed that
the action of symmetries of the ``first kind" on $\tau$-functions
is that of a differential operator with coefficients polynomially depending on
the time variables). Van Moerbeke gave another proof [2] which was more direct
and free of that conjecture.

The main goal of this note is to give a new, and possibly short, proof to
this important result (Theorem 2 below). The proof is based on a new expression
for a generator of additional symmetries (Theorem 1). The formula is very
simple and good looking, it generalizes an earlier obtained expression (see
[3]) for resolvents, generators of symmetries belonging to the hierarchy.
It was our next goal to present this formula.\\

{\bf 1.} Here we summarize well-known definitions and properties of the
KP hierarchy.

The KP hierarchy is generated by a pseudodifferential operator ($\Psi
$DO)$$L=\p+u_1\p^{-1}+u_2\p^{-2}+...,~~\p=d/dx.$$ This operator can be
represented in a dressing form $$L=\f\p\f^{-1}$$where $\f$ is a $\Psi$DO
$\f=\sum_0^\infty w_i\p^{-i}$ with $w_0=1$. Assuming that $w_i$ depend on some
``time variables" $t_m$, where $m=1,2,... $ the hierarchy is the totality of
equations $$\p_m\f=-L_-^m\f,~~\p_m=\p/\p t_m.$$ Let $\xi(t,z)=\sum_1^\infty
t_iz^i.$ Put $$w(t,z)=\f\exp\xi(t,z)=\sum_0^\infty w_iz^i\exp\xi(t,z)=\hw(t,z)
\exp\xi(t,z).$$This is the (formal) {\em Baker, or wave, function}.

Let $\f^*$ be the formal adjoint to $\f$ (by definition, $(f\p)^*=-\p f$).
The function $w^*(t,z)=(\f^*)^{-1}\exp(-\xi(t,z))=\hw^*(t,z)\exp(-\xi(t,z))$ is
called the {\em adjoint Baker function}.

Further we need a very simple and useful lemma (see [3]).
One can consider two types of residues, that of formal series in $\p$:
res$_{\p}\sum a_i\p^i=a_{-1}$, and that of formal series in $z$: res$_z\sum
a_iz^i=a_{-1}$.\\

{\bf Lemma 1.} {\sl Let $P$ and $Q$ be two $\Psi$DO,
then $${\rm res}_z~[(Pe^{xz})\cdot(Qe^{-xz})]={\rm res}_{\p}~PQ^*$$ where $Q^*$
is the formal adjoint to $Q$.}\\

The proof is in a straightforward verification. We also need another
simple lemma:\\

{\bf Lemma 2.} {\sl If $f(z)=\sum_{-\infty}^\infty a_iz^{-i}$ then $${\rm res}_
z[\z^{-1}(1-z/\z)^{-1}+z^{-1}(1-\z/z)^{-1}]f(z)=f(\z).$$ (Here $(1-z/\z)^{-1}$
is understood as a series in $\z^{-1}$ while $(1-\z/z)^{-1}$ is a series in
$z^{-1}$).}\\

The Baker function can be expressed in terms of the $\tau$-function as
$$\hw(t,z)={\tau(t_1-1/z,t_2-1/(2z^2),t_3-1/(3z^3),...)\over\tau(t_1,t_2,t_3,
...)}={\tau(t-[z^{-1}])\over\tau(t)}.$$A $\tau$-function is determined
up to a multiplication by $c\exp\sum_1^\infty c_it_i$ where $c,c_1,c_2,...$
are arbitrary constants. For the adjoint Baker function we have
$$\hw^*(t,z)={\tau(t_1+1/z,t_2+1/(2z^2),t_3+1/(3z^3),...)\over\tau(t_1,t_2,t_3,
...)}={\tau(t+[z^{-1}])\over\tau(t)}.$$

{\bf 2.} The above formulas can be rewritten in terms of the co-called vertex
operators. One can write $f(t-[z^{-1}])=\exp(-\sum_1^\infty\p_i/(iz^i))f(t)$.
Then $$\hw(t,z)={\exp(\sum_1^\infty t_iz^i)\exp(-\sum_1^\infty\p_i/(iz^i))\tau(
t)\over\tau(t)}.$$ Moreover, for two non-commuting operators $A$ and $B$
an equation can be written $$e^A\cdot e^B=:e^{A+B}:$$ where the symbol of
normal
ordering ``: :" means that in all monomials the operator $A$ must be placed to
the left of all $B$'s.

Let $$p_i=\left\{\begin{array}{ll}\p_i,&i>0\\|i|t_{|i|},&i\leq 0\end{array}
\right.$$ These are Heisenberg generators. Then the above formula for $\hw$ is
$$\hw(t,z)={:\exp(-\sum_{-\infty}^\infty p_i/(iz^i)):\tau(t)\over\tau(t)}.$$
The symbol of normal ordering means here that $p_i$ with negative $i$ must
be placed to the left of positive ones. The operator $X(z)=
:\exp\sum_{-\infty}^\infty p_i/(iz^i):$ is called a vertex operator. Similarly,
$$\hw^*(t,z)={:\exp\sum_{-\infty}^\infty p_i/(iz^i):\tau(t)\over\tau(t)}.$$

Another vertex operator can be introduced:
$$X(\l,\m)=:\exp\sum_{-\infty}^\infty
({p_i\over i\l^i}-{p_i\over i\m^i}):.$$

{\bf Proposition 1 (Sato).} {\sl The operator $X(\l,\m)$ acts as an
infinitesimal operator
in the space of $\tau$-functions, i.e., solving the differential equation
$\p\tau/\p t_{\l,\m}^*=X(\l,\m)\tau$ where $t_{\l,\m}^*$ is a variable, we
obtain for each $t_{\l,\m}^*$ a new $\tau$ function. In other words this yields
symmetries of the KP hierarchy.}\\

The operator $X(\l,\m)$ can be considered as a generator of infinitesimal
symmetries if expanded in double series, in $\m-\l$ and $\l$:
$$X(\l,\m)\tau=\sum_{m=0}^\infty {(\m-\l)^m\over m!}\sum_{n=-\infty}^\infty\l^
{-n-m}W_n^{(m)}(\tau).$$ Differential operators $W_n^{(m)}$ can be taken as
generators of a Lie algebra which is called $W_{1+\infty}$.

The procedure of application of $X(\l,\m)$ to $\tau$ is called Sato's
(infinitesimal) B\"acklund transformation. \\

{\bf 3.} Now we define additional symmetries in a form given them by Orlov
and Shulman [4] (see also [3]). Let $$\Gamma=\sum_1^\infty t_ii\p^{i-1}
{}~~{\rm and}~~~M=\f\Gamma\f^{-1}.$$

{\bf Proposition 2 (Orlov and Shulman).} {\sl The differential equation
$$\p^*_{lm}\f=-(M^mL^l)_-\f$$ where $\p^*_{lm}$ symbolizes a derivative
with respect to some additional variable $t^*_{lm}$ gives a symmetry of the
KP hierarchy.}\\

One can consider a generator of these symmetries
$$Y(\l,\m)=\sum_{m=0}^\infty{(\m-\l)^m\over m!}\sum_{l=-\infty}^\infty \l^
{-m-l-1}(M^mL^{m+l})_-.$$

{\bf Theorem 1.} {\sl The formula $$Y(\l,\m)=w(t,\m)\cdot\p^{-1}\cdot w^*(t,\l)
$$ holds.}\\

{\bf Remark.} {\em If $\l=\m$ then $Y(\l,\l)=\sum_{-\infty}^\infty \l^{-l-1}
L_-^l$ is the so-called resolvent which is a generator of the symmetries
belonging to the hierarchy; in this special case the theorem 1 was
proven in [3].}\\

{\em Proof of the theorem.} We have $$(M^mL^{m+l})_-=(\f\G^m\p^{m+l}\f^{-1})_-=
\sum_1^\infty \p^{-i}{\rm res}_\p~\p^{i-1}\f\G^m\p^{m+l}\f^{-1}.$$ According to
Lemma 1 this can be written as $$(M^mL^{m+l})_-=\sum_1^\infty \p^{-i}{\rm res}_
z~\p^{i-1}\f\G^m\p^{m+l}e^{\xi(t,z)}(\f^*)^{-1}e^{-\xi(t,z)}.$$ Taking into
account that $$\G\exp\xi(t,z)=\sum_1^\infty t_ii\p^{i-1}\exp\xi(t,z)=
\sum_1^\infty t_iiz^{i-1}\exp\xi(t,z)=\p_z\exp\xi(t,z)$$ and that $\f$ commutes
with $\p_z$ we have
$$(M^mL^{m+l})_-={\rm res}_z~\sum_1^\infty \p^{-i}(z^{m+l}\p_z^mw)^{(i-1)}
\cdot w^*=\res_z~z^{m+l}\p_z^mw\cdot\p^{-1}\cdot w^*.$$Now, using Lemma 2,
we have $$Y(\l,\m)={\rm res_z}~\sum_{m=0}^\infty\sum_{l=-\infty}^\infty{z^{m+l}
\over \l^{m+l+1}}\cdot
{1\over m!}(\m-\l)^m\p_z^mw\cdot\p^{-1}\cdot w^*$$ $$={\rm res_z}~[{1
\over z(1-\l/z)}+{1\over \l(1-z/\l)}]\exp((\m-\l)\p_z)w(t,z)\cdot \p^{-1}
\cdot w^*(t,z)$$ $$=\exp((\m-\l)\p_\l)w(t,\l)\cdot\p^{-1}\cdot w^*(t,\l)
=w(t,\m)\cdot\p^{-1}\cdot w^*(t,\l).~~\Box$$               \\

{\bf 4.} Our second result is a new proof of the following theorem:\\

{\bf Theorem 2 (Adler, Shiota, van Moerbeke).} {\sl The action of the
infinitesimal operator $X(\l,\m)$ on the Baker function $w(t,z)$ generated by
its action on $\tau$ given by Sato's formula is connected with $Y(\l,\m)$ by
the formula} $$X(\l,\m)=(\l-\m)Y(\l,\m).$$

This means that the additional symmetries in Fuchssteiner - Chen, Lee and Lin -
Orlov and Shulman sense are the same as Sato's B\"acklund transformations.

{\em Proof.} We need two identities involving the $\tau$-function, the Fay
identity and its differential form, see, e.g., [5].\\

{\bf Fay identity.} {\sl The $\tau$-function satisfies the identity
$$\sum_{(s_1,s_2,s_3)}(s_0-s_1)(s_2-s_3)\tau(t+[s_0]+[s_1])\tau(t+[s_2]+[s_3])
=0$$ where $(s_1,s_2,s_3)$ symbolizes cyclic permutations.}\\

{\bf Differential Fay identity.}$$\p\tau(t-[s_1])\cdot\tau(t-[s_2])-
\tau(t-[s_1])\cdot\p\tau(t-[s_2])$$
$$+(s_1^{-1}-s_2^{-1})\{\tau(t-[s_1])\tau(t-
[s_2])-\tau(t)\tau(t-[s_1]-[s_2])\}.$$

We have
$$X(\l,\m){\tau(t-[z^{-1}])\over\tau(t)}={\tau(t)X\tau(t-[z^{-1}])-\tau(t-
[z^{-1}])X\tau(t)\over\tau^2(t)}$$ $$=\{[\tau(t)\exp\xi(t,-\l+\m)(1-{\l\over z}
)^{-1}(1-{\m\over z})\tau(t-[z^{-1}]+[\l^{-1}]-[\m^{-1}])$$ $$-\exp\xi(t,-\l+\m
)\tau(t-[z^{-1}])\tau(t+[\l^{-1}]-[\m^{-1}])\}/\tau^2(t)$$ $$=\exp\xi(t,-\l+\m)
(z-\l)^{-1}\{t'+[z^{-1}]+[\m^{-1}])\tau(t'+[\l^{-1}])(z-\m)$$ $$-\tau(t'+[\m^{-
1}])\tau(t'+[\l^{-1}]+[z^{-1}])(z-\l)\}/\tau^2(t)$$ where $t'=t-[z^{-1}]-[\m^{-
1}]$. The expression in the braces can be transformed according to the Fay
identity ($s_0=0,s_1=\l^{-1},s_2=\m^{-1},s_3=z^{-1}$). It is equal to
$-(\m-\l)\tau(t'+[z^{-1}])\tau(t'+[\l^{-1}]+[\m^{-1}])$. In order to obtain the
action of X on $w(t,z)$ we must multiply this by $\exp\xi(t,z)$.

Now, the equality $$\exp\xi(t,z)\exp\xi(t,-\l+\m)(z-\l)^{-1}(\l-\m)\tau(t-
[\m^{-1}])\tau(t+[\l^{-1}]-[z^{-1}])/\tau^2(t)$$
$$=(\l-\m)w(t,\m)\p^{-1}w^*(t,\l)\hw(t,z)\exp\xi(t,z) $$ must be proven.
Dividing by $(\l-\m)w(t,\m)$ and multiplying by $\p$ we have
$$\p\exp\xi(t,z-\l)(z-\l)^{-1}{\tau(t+[\l^{-1}]-[z^{-1}])\over\tau(t)}$$ $$=
{\tau(t+[\l^{-1}])\tau(t-[z^{-1}])\over\tau^2(t)}\exp\xi(t,z-\l)$$ or
$$\p\tau(t+[\l^{-1}]-[z^{-1}])\cdot\tau(t)-\tau(t+[\l^{-1}]-[z^{-1}])\cdot
\p\tau(t)$$ $$=-(z-\l)\{\tau(t+[\l^{-1}]-[z^{-1}])\tau(t)-\tau(t+[\l^{-1}])\tau
(t-[z^{-1}])\}$$ which is the differential Fay identity, i.e. it is true. All
the transformations are convertible. $\Box$\\

{\bf 5.} We are going to present one more formula for Sato's generator
$X(\l,\m)$ which can be useful. This formula appeared as a result of
discussions with M. Niedermaier and, virtually, belongs to him. We merely
simplified the proof. Let $$X(\l,\m)\tau=\sum_{m=0}^\infty ((\m-\l)^m/m!)W^{(m)
}(\l),~~~W^{(m)}(\l)=\sum_{n=-\infty}^\infty\l^{-n-m}W_n^{(m)}.$$  Let
$$\theta(\l)=\sum_{-\infty}^\infty
{p_i\over i\l^i},~{\rm then}~~X(\l,\m)=~:\exp(\theta(\l)-\theta(\m)):~.$$
The normal ordering means that we can
operate with all operators as if they commute. Then  $$X(\lambda,\mu)=~
:e^{\theta(\lambda)}e^{-\theta(\mu)}:~=~:[\sum_m{(\mu-\lambda)^m\over m!}
\p_\mu^me^{-\theta(\mu)}|_{\mu=\lambda}]e^{\theta(\lambda)}:~.$$ Now,
$$W^{(m)}(\lambda)=~:\p_\lambda^me^{-\theta(\lambda)}\cdot e^{\theta(\lambda)}:
{}~=Q_m(\lambda).$$ Polynomials $Q_m(\lambda)$ satisfy the recursion relations
$$Q_0(\lambda)=1,~Q_{m+1}(\lambda)=(\p_\lambda-\theta'(\lambda))Q_m(\lambda).$$
They are $Q_m(\lambda)=P_m(-\theta'(\lambda))$ where $P_m$ are the so-called
Fa\`a di Bruno polynomials (see, e.g., [3]). It is easy to prove by induction
using the above recursion formula that $$W^{(m)}(\l)=\sum_{m_1+2m_2+...+km_k=m}
:{m!\over m_1!m_2!...m_k!}(-\p_\l\theta/1!)^{m_1}(-\p_\l^2\theta/2!)^{m_2}...
(-\p_\l^k\theta/k!)^{m_k}:~$$ (see [3], Eq.(7.5.7)).\\

{\bf References.}\\

\noindent 1. Adler M., Shiota T., and van Moerbeke P., From the
$w_\infty$-algebra to its central extension: a $\tau$-function approach,
to appear.\\

\noindent 2. Van Moerbeke P., Lecture in Colloquium on Integrable Systems
and Topological Field Theories, Cortona, Italy, September 1993.\\

\noindent 3. Dickey L. A., Soliton equations and Hamiltonian systems, Adv.
Series in Math. Phys., vol. 12, World Scientific, 1991.\\

\noindent 4. Orlov A. Yu. and Shulman E. I., Additional symmetries for
integrable and conformal algebra representation,
Lett. Math. Phys., 12, 171, 1986\\

\noindent 5. Adler M. and van Moerbeke P., A matrix integral solution to
two-dimensional $W_p$-Gravity, CMP, 147, 25-56, 1992\\

\end{document}